\documentclass[aps,floats,preprint,superscriptaddress]{revtex4-1}

\renewcommand{\thesection}{\arabic{section}}

\usepackage{amssymb}
\usepackage{amsmath}
\usepackage{graphicx}
\usepackage{dcolumn}
\usepackage{color}
\usepackage{amsmath}

\begin{document}

\title{Strategical incoherence regulates cooperation in social dilemmas on multiplex networks}

\author{Joan T. Matamalas}
\affiliation{Departament d'Enginyeria Inform\`atica i Matem\`atiques, Universitat Rovira i Virgili, 43007, Tarragona, Spain.}

\author{Julia Poncela-Casasnovas}
\affiliation{Departament d'Enginyeria Inform\`atica i Matem\`atiques, Universitat Rovira i Virgili, 43007, Tarragona, Spain.}

\author{Sergio G\'omez}
\affiliation{Departament d'Enginyeria Inform\`atica i Matem\`atiques, Universitat Rovira i Virgili, 43007, Tarragona, Spain.}

\author{Alex Arenas}
\affiliation{Departament d'Enginyeria Inform\`atica i Matem\`atiques, Universitat Rovira i Virgili, 43007, Tarragona, Spain.}

\begin{abstract}

Cooperation is a very common, yet not fully-understood phenomenon in natural and human systems. The introduction of a network within the population is known to affect the outcome of cooperative dynamics, allowing for the survival of cooperation in adverse scenarios. Recently, the introduction of multiplex networks has yet again modified the expectations for the outcome of the Prisoner's Dilemma game, compared to the monoplex case. However, much remains unstudied regarding other social dilemmas on multiplex, as well as the unexplored microscopic underpinnings of it.
In this paper, we systematically study the evolution of cooperation in all four games in the $T-S$ plane on multiplex. More importantly, we find some remarkable and previously unknown features in the microscopic organization of the strategies, that are responsible for the important differences between cooperative dynamics in monoplex and multiplex. Specifically, we find that in the stationary state, there are individuals that play the same strategy in all layers (coherent), and others that don't (incoherent). This second group of players is responsible for the surprising fact of a non full-cooperation in the Harmony Game on multiplex, never observed before, as well as a higher-than-expected cooperation rates in some regions of the other three social dilemmas.

\end{abstract}

\maketitle
\section{Introduction}
\label{sec:introduction}
Cooperation is a ubiquitous and yet not fully-understood phenomenon in Nature: from humans that cooperate to build complex societies to animals like wolves that hunt in packs in order to catch preys larger than they are, or meerkats that watch out for predators in turn while the rest of the colony feeds. Even small microorganism cooperate to survive in hostile environments. For instance, the {\em Dictyostelium discoideumu}, usually a solitary amoeba, when starves it associates with others in order to form a multicellular slug for the sake of survival. Explaining how cooperation has emerged and has resisted against more selfish behaviours is one of the biggest challenges in natural and social sciences. From a mathematical point of view, the problem of cooperation within a population can be studied using Evolutionary Game Theory~\cite{Smith:1982ww,Szabo:2007uy,Anonymous:D964cQ76}. There are multiple mechanisms proposed to explain the evolution of cooperation, such as reputation, kin selection, network reciprocity or punishment~\cite{Nowak:2006bt,Pennisi:2005hx,Hauert:2007js}. On the other hand, outstanding experimental efforts have been made in the last few years~\cite{Grujic:2010jm,GraciaLazaro:2012jp,10.1371/journal.pone.0016836,Rand:2011dv,Grujic:2014dr} to try to understand how actual humans behave when confronted with social dilemmas in a formal Game Theory environment.

We focus here on the impact of the structure of the network of interactions among individuals on the outcomes of the cooperation dynamics. The study of networks, their properties and dynamics, has experimented a huge advance in the last few decades, empowered by the technological advances that enable the acquisition of real data about interactions between individuals from social networks~\cite{Wellman:1988ve,Wasserman:1994wt}, mobile communication networks~\cite{gonzalez:2008hya} or collaborations between scientific authors~\cite{Newman:2004fw}. There is a vast literature on the evolution of cooperation on complex networks~\cite{Santos:2005ur, perc2010coevolutionary, perc2013evolutionary}, studying aspects ranging from the effect of network topology on cooperation~\cite{GomezGardenes:2006fi} to network growth driven by cooperation dynamics~\cite{Poncela:2008cc,Poncela:2012tk}, and other spatial and temporal effects~\cite{Roca:2009joa} that offer insights on how cooperation can evolve and survive in different scenarios.

\begin{figure*}[t!]
\centering
\includegraphics[width=0.20\textwidth]{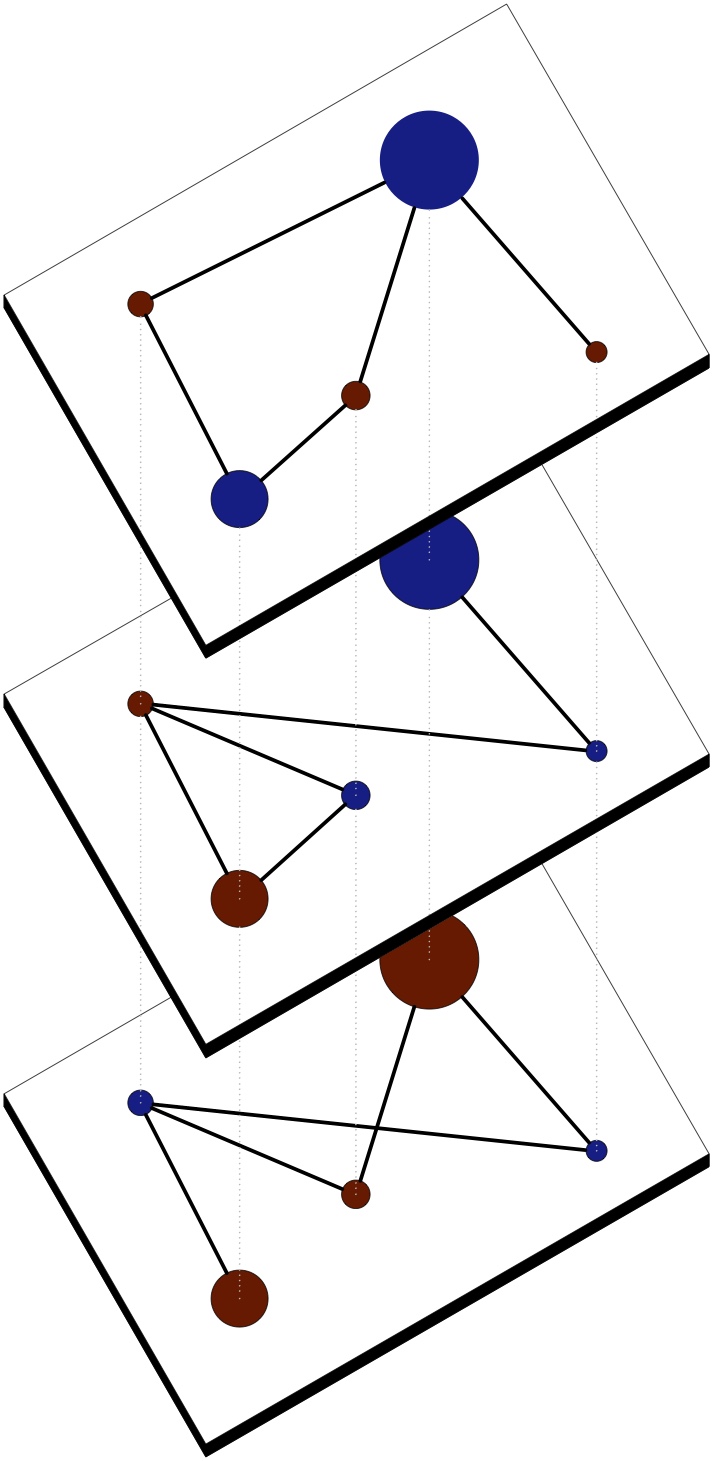}
\caption{Example of a multiplex network with 3 layers, 5 nodes per layer and 5 links in each layer. The color of the nodes represents the strategy played in that layer, red for cooperators, blue for defectors. Their size is proportional to their global payoff.}
\label{fig:multiplex}
\end{figure*}

An innovative way of representing multiple types of social interactions in one single structure is the use of multiplex networks~\cite{Mucha:2010bz,Kurant:2006dm,Kivela:2013wm,DeDomenico:2013ef}, see Fig.~\ref{fig:multiplex}, which have been already successfully applied to the study of disease spreading~\cite{Granell:2013hy} and diffusion dynamics~\cite{Gomez:2012if} (for a complete review look at~\cite{Boccaletti:2014bz}). Multiplex networks are interesting in this field, because many social interactions can be understood as a combination of interactions at different, independent levels, each one representing a different social scenario such as family, friends, coworkers, etc. An individual's behaviour can be different in each level, but it is ultimately conditioned by all of them. Some work has been done to understand the evolution of the Prisoner's Dilemma game on multiplex networks~\cite{GomezGardenes:2012hca}, exploring different coupled evolutionary games using a interdependent networks~\cite{Santos:2014}.
The impact of the degree correlations among layers~\cite{Anonymous:QiIVHrUI} on the outcome of social dilemmas have also been studied on 2-layer network, where one layer was used for the accumulation of payoffs and the other for strategy updating.
There are also works that explore the problem of cooperation on coupled networks~\cite{Wang:2012ex}, and even optimizing the interdependence between them via coevolution~\cite{Wang:2014so}.
However, the evolution of cooperation on top of multiplex networks with any number of layers hasn't been systematically studied for all four social dilemmas.

The objective of this paper is, on the one hand, to provide an exhaustive analysis of the resilience and propagation of cooperation in the main four social dilemmas in Game Theory literature, studying the average levels of cooperation, payoff distribution, and dependence on the initial fraction of cooperation, as a function of the number of layers of the multiplex. More importantly, we will focus on analyzing the previously unexplored microscopic behaviour of individuals across layers.

This work is organized as follows. In Section~\ref{sec:model} we define the model we have used in this work. Section~\ref{sec:results} contains our findings on the density of cooperators for each one of the proposed scenarios. Then we turn our attention to the microscopic behaviour of individuals across different layers. Finally, a summary and conclusions can be found in Section~\ref{sec:discussion}.

\section{Model}
\label{sec:model}
We will focus on two-strategy social dilemmas. If we assume that each player in the system can either cooperate (C) or defect (D), a game can be defined according to its payoff matrix:
\begin{equation}
\bordermatrix{
  & C & D \cr
C & R & S \cr
D & T & P \cr}
\label{tab:payoffMatrix}
\end{equation}
Where \emph{R} represents the reward obtained by a cooperator playing against another cooperator, \emph{S} is the sucker payoff obtained by a cooperator when it plays against a defector, the temptation payoff, \emph{T}, is the payoff received by a defector when his opponent is a cooperator, and finally, \emph{P} represents the payoff obtained by a defectors which engages with another defector.

Traditionally the values of \emph{R} and \emph{P} are fixed to $R=1$ and $P=0$ in order to provide a fixed scale for the game payoffs~\cite{Nowak:1992vx, Hauert:2004jm}. Applying this constraint, it turns out that the selection of the remaining parameters T and S enables the definition of several games according to their evolutionary stability. Thus, if $R > S > P$ and $R > T > P$ the game is the harmony game~\cite{Licht:1999vr}. The final state of a population playing this game will be total cooperation, regardless of the initial fraction of cooperators. Prisoner's dilemma~\cite{Axelrod:1980vr, Axelrod:1981ba, Rapoport:1974jp}, $T>R>P>S$, represents the opposite situation, and the population evolves towards total defection regardless of the initial conditions (although all players would be better off cooperating, hence the dilemma). A classical example of a coordination game, the stag-hunt game~\cite{Rousseau:2003vj, Luce:2012un}, is represented when the payoff values respect the order $R>T>P>S$, the output of this game will be either total defection or total cooperation, depending on the initial conditions. Finally, an anti-coordination game, the Hawk-Dove~\cite{Smith:1982ww,Smith:1976tz}, takes place if the payoff values follows $T > R > S > P$, where the final state will be a population made of both cooperators and defectors.

The players sit on the nodes of a multiplex network of $L$ layers. Each node is present in all layers, but in general, they have different connectivity in each layer. Every layer, $\ell_i$, in the multiplex network is a connected and homogeneous Erd{\H{o}}s-R\'{e}nyi  (ER) network, with the same number of edges $E$ and nodes $N$, and equal degree distribution, and the multiplex network is generated avoiding degree correlations between layers. Each layer is represented by an adjacency matrix $A^{\ell}$, where $A^\ell_{ij} = 1$ if nodes $i$ and $j$ are connected in that layer, and $A^\ell_{ij} = 0$ otherwise. That representation enables the definition of the degree of node $i$ in layer $l$ as $k_i^{\ell} = \sum_{j=1}^{N}{A_{ij}^{\ell}}$ and its global degree in the multiplex as $K_i = \sum_{\ell=1}^{L}{k_i^{\ell}}$.

Each round of the game is divided in two phases: payoff recollection and strategy update. Each node $i$ can choose to play one of the two strategies, cooperation or defection, independently in each layer of the network and at every time step, $s^\ell_i(t)$. Within a specific payoff matrix, the node $i$'s strategy determines the payoff, $p_i^{\ell}$, that it obtains in a layer $l$ when it plays against all its $k_i^{\ell}$ neighbors. The total payoff of node $i$ can be easily calculated as $P_i = \sum_{\ell=0}^{L}{p_i^{\ell}}$. At the end of each round, each player can change the strategy in one of its layers, $s_i^{\ell}$, using the Replicator-like Rule: A node chooses a layer of the multiplex, $\ell_r$, with uniform probability. Then it chooses with uniform probability one of its $k_i^{\ell_r}$ neighbors, $j_r$, in that layer. If $P_i < P_{j_r}$ and $s_i^{\ell} \neq s_{j_r}^{\ell}$ the probability that node $i$ changes its strategy in layer $\ell_r$ is given by:
\begin{equation}
	\Pi_{i\rightarrow j_r}^{\ell_r}{(t)} = \frac{P_{j_r}(t) - P_{i}(t)}{\max(K_i,K_{j_r})\cdot (\max(1,T) - \min(0,S))}
    \label{eq:update_rule}
\end{equation}

It is important to notice that the update rule uses global information about the players: global degree and global payoff (that is, added up over all layers), in order to update the strategy of any particular layer. That is the way our model shares information between layers and relies in the social nature of layers' interdependency~\cite{GomezGardenes:2012hca}: each player only has information about the strategy of its neighbour in their same layer (but not in those layers where they are not connected). However, it knows its neighbor's total benefits, and it makes the simplifying assumption that it is using the same strategy in every layer. As we will see later on, this fact has a profound impact on the outcomes of the dynamics, compared to the monoplex scenario.

At the end of each time step the density of cooperators can be computed for each layer and for the entire multiplex using:

\begin{equation}
    c(t) = \frac{1}{L}\sum_{\ell=1}^{L}{c^{\ell}(t)} = \frac{1}{L \cdot N}\sum_{\ell=1}^{L}{\sum_{i=1}^{N}{s^\ell_i(t)}}
    \label{eq:level_of_cooperators}
\end{equation}

\section{Results}
\label{sec:results}

To ascertain the outcome of the cooperative dynamics for the different games on multiplex networks, we will start by studying the stationary level of cooperation in the system, then we will study the effect of the initial fraction of cooperators, and finally, we will move to analyzing in detail the microscopic organization of cooperation for individuals across different layers.

The results are obtained for a range of values of $T\in[0,2]$ and $S\in [-1,1]$ that defines the $T-S$ plane. The simulation runs on a multiplex network that has $N=1000$ nodes and $E=3000$ edges per layer distributed according an Erd{\H{o}}s-R\'{e}nyi degree distribution with $\langle k \rangle$. For each possible pair of values of the game parameters the simulation runs $1\times 10^5$ time steps, that is the transient time $t_0$ needed by the algorithm to generally reach a stationary state (we further discuss the matter of convergence time in the Supplementary Information, Section~\ref{sec:convergence}). After this time the algorithm runs for another $t_{\gamma}=2\times 10^4$ time steps. All the quantities of interest are averaged over this second period of time. The experiments are repeated and averaged over  $I=64$ different networks and initializations in order to gain statistical confidence. The initial fraction of cooperators, $c_0$, is distributed randomly in each layer. We focus here on the case $c_0=0.5$, although we have also explored other values (see Supplementary Information, Section~\ref{sec:initial_conditions}).

\begin{figure*}[t!]
    \centering
    \includegraphics[width=0.95\textwidth]{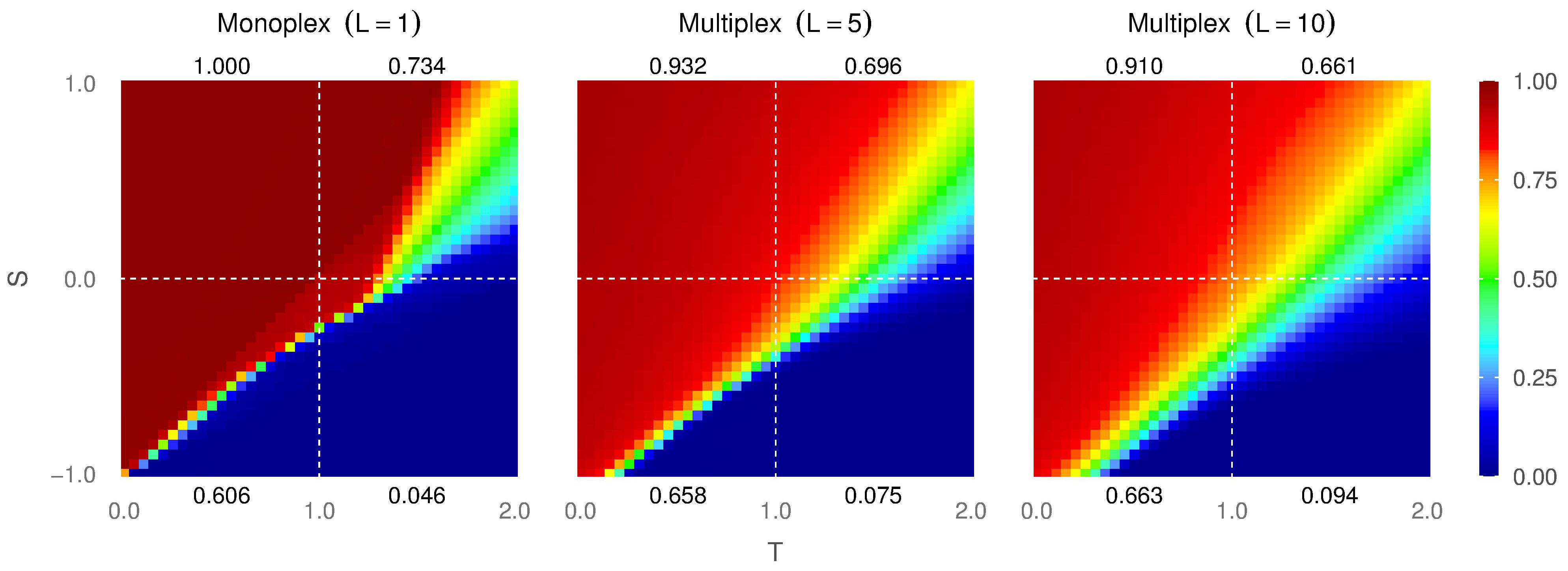}
    \caption{Asymptotic density of cooperators $\langle c\rangle$ for networks with different number of layers ($L=1$ on the left, $L=5$ in the middle, $L=10$ on the right). The plane $T-S$ is divided into four major regions that correspond to the four games under study: the upper-left area is the Harmony Game, the upper-right is the Snow Drift, Stag-Hunt is in the lower-left, and the Prisoner's Dilemma is in the lower-right. The average asymptotic density of cooperators for each one of the games is also indicated, as a numerical value, next to the corresponding quadrant. See Supplementary Information, Section~\ref{sec:initial_conditions}, for the corresponding results for other values of the initial fractions of cooperation.}
\label{fig:average}
\end{figure*}

\textbf{Density of cooperators.}
The stationary average value of cooperation is defined according to the following:

\begin{equation}
    \langle c \rangle = \frac{1}{t_{\gamma}\cdot I}\cdot \sum_{i=1}^{I}{\sum_{t=t_0}^{t_0+t_{\gamma}}{c_i(t)}}
    \label{eq:asymptoticDensity}
\end{equation}

In Fig.~\ref{fig:average} we present the average stationary value of cooperation when $c_0 =0.5$. We observe that our results for the monoplex case (left) are consistent with those obtained by Roca et al.~\cite{Roca:2009joa} for this kind of networks. The results for multiplex networks show a large increase of the areas where both strategies coexist (that is, the areas in the plane that separate total cooperation from total defection). However, this coexistence is of a different nature depending on the evolutionary stability of the particular game (or quadrant), as we explain below.

The Stag Hunt game has an unstable evolutionary equilibrium with mixed populations. This means that, when there is a structure, the population will evolve towards total cooperation or total defection depending on the initial population and type of structure of the network (due to this fact, the standard deviation of the $\langle c \rangle$ is large in that transition area, see Supplementary Information, Section~\ref{sec:convergence}, for details). For the monoplex we have a very narrow transition area between total cooperation or defection populations (left panel in Fig.~\ref{fig:average}). This transition region widens with the number of layers, enabling the coexistent of both strategies in a larger portion of the game parameter space. The explanation of such behaviour can be found in the inter-layer dynamics: it is more likely that a cooperator or a defector node resists in hostile environments in a particular layer, because its fitness is not evaluated in just that layer, but also in the other layers where, due to its strategy or its topological  configuration, the node might have better performance. The Stag Hunt game, where the maximum payoff possible is obtained when a cooperator plays against another cooperator, favors specially the resilience of cooperators nodes when the temptation value is low: a cooperator node $i$ in layer $\ell_r$ that has a big payoff $P_i$ has higher probability of spreading its strategy to its defector neighbours in $\ell_r$, thus increasing its payoff. This increase will propagate to the other layers, making the strategies of the player more robust against invasion. Playing defection in layer $\ell_r$ when temptation value is small, does not have a big effect in the global payoff of the node. As a consequence, in this particular game the multiplex structure increases specially the resilience of cooperators, thus the average density of cooperators in this game quadrant shows an statistically significant increase as we keep adding layers to the structure (Mann-Whitnet U test, $\alpha = 0.05$).

In the Prisoner's dilemma game, defection dominates cooperation. Related papers~\cite{Roca:2009joa} show that for ER networks using Replicator rule when temptation and sucker payoffs are not too large, cooperation can survive forming groups of cooperative clusters, thus resisting against the initial attempt of invasion by defectors, and then spread through the population. Our results for the monoplex are consistent with those. For the multiplex, we observe how the transition region between all-cooperator and all-defector situations is larger than for the monoplex, as in the case of Stag Hunt game. It is worth noticing that regions where we have all-cooperator populations in the monoplex, are not necessarily all-cooperator areas in its multiplex counterpart. This happens because the formation of cooperative clusters in one layer will also increase the fitness of these nodes in the other layers regardless of the strategy used in these other layers. And this can lead to a reinforcement of defector strategies due to the inter-layer dynamics, increasing their survival rate. This inter-layer dynamics will led to a widening of the transition area that enables survival of cooperators in areas where they are not present in the monoplex scenario. If we take into account the whole Prisoner's Dilemma quadrant, the conclusions are the same that in the Stag Hunt game: a statistically significant increase in the average density of cooperators occurs as we increase the number of layers.

The Snow Drift game has a stable equilibrium in mixed populations: it is an anti-coordination game. Previous works~\cite{Roca:2009joa} show that for ER networks there are some regions in the plane $T-S$ for which this game converges to single-strategy populations. For lower values of the temptation these regions are prone to cooperation. In multiplex networks however, single strategy regions are less common and mixed populations are the rule. That happens by the same inter-layer dynamics that we have explained earlier: the impact of a cooperator's benefits on the other layers of the multiplex structure. This entails a significant reduction on the average fraction of cooperators from $0.734$ in the monoplex to $0.661$ in the 10-layer multiplex for this quadrant.

Finally, the Harmony game has cooperation as its dominant strategy. For  single-layer ER networks with Replicator update rule, Roca et al.~\cite{Roca:2009joa} reported that the whole quadrant ends up in an all-cooperator configuration. However, in the case of multiplex scenarios, the average fraction of cooperators decreases significantly as we keep adding layers to the system: $0.932$ for $L=5$ and $0.910$ for $L=10$. This increasing resilience of defection can be explained as a consequence of the multiplex topology and the lack of degree correlations between layers: due to the payoff accumulated by an individual acting as cooperator in some layers, defector nodes can resist against cooperators in other layers.

We can mathematically prove that defectors can survive and be stable in the Harmony game on ER multiplex networks by analyzing the simplest situation: let's  assume a multiplex structure with $L$ layers. In one single layer (for simplicity we assume it will be the first one) we have one single node playing as defector, but it plays as cooperator in all the other $L-1$ layers. There are no more defectors anywhere in the system. This node's connectivity in layer $\alpha$ is $k_\alpha$, and, recalling that $R=1$ and $P=0$, the total payoff of that node that is defecting in one single layer is given by:

\begin{equation}
    P_d  = T k_1 + \sum_{\alpha =2}^{L}k_{\alpha}
    \label{eq:payoff_D}
\end{equation}

The payoff of any of the node's neighbors (note that all of them play as cooperators), with a degree $k'_\alpha$ in layer $\alpha$, is:

\begin{equation}
    P_c  =  (k'_1 - 1) + S + \sum_{\alpha=2}^{L}k'_{\alpha} = \sum_{j=1}^{L}k'_{\alpha} + S -1
    \label{eq:payoff_C}
\end{equation}

Thus, in order to survive as a defector in layer $\alpha$, the following inequality must be fulfilled for each of the node's neighbours:

\begin{align}
    P_d  &\geq  P_c \\
    T k_1 + \sum_{i=2}^{L}k_i   &\geq   \sum_{j=1}^{L}k'_j + S -1
    \label{eq:condition_no_invasion}
\end{align}

We can estimate both a soft and a hard limit for the previous inequality. As a soft limit, and assuming we have independent, uncorrelated Erd{\H{o}}s-R\'{e}nyi layers in our multiplex network, we can approximate every $k_{\alpha}$ by $\langle k \rangle$ and get:

\begin{align}
    (T +L - 1)\langle k \rangle   \geq   S-1+L\langle k \rangle \\
    \langle k \rangle   \geq  \frac {S-1}{T - 1}
    \label{eq:soft_limit}
\end{align}

On the one hand, a hard limit for the condition can be calculated by approximating $k_i$ by $k_{\max}$ for the cooperator neighbours:

\begin{align}
    (T +L-1) \langle k \rangle   &\geq   S-1 + L k_{\max}  \\
    \langle k \rangle   &\geq  \frac {S-1 + L k_{\max} }{T +L - 1}
    \label{eq:hard_limit}
\end{align}

On the other hand, we can calculate the probability of this topological situation happening. First of all we have to define what is the probability of a node $i$ to have degree $k$, $P(X=k)$. In our model, and in order to avoid the non-negligible effect of unconnected nodes, we impose a minimum connectivity, $k_{\min}$. To get a more accurate approximation of our degree distribution we take into account this minimum:

\begin{equation}
P_{k_{\min}}(X=k) = \frac{P(X=k)}{1-P(X\le k_{\min})}
\label{eq:probKmin}
\end{equation}

As it has been stated previously, the payoff of cooperators against cooperators is proportional to their degree, since we set $R=1$: in this example we use $L-1$ full cooperative layers, so the payoff obtained in this layers is proportional to the degree distribution of the aggregate network of this $L-1$ layers. Moreover, the payoff distribution of the nodes that play cooperation in all layers is proportional to the aggregation of all layers, $L$. Imposing that we do not have inter-layer degree correlation, the degree distribution of the aggregated networks can be modeled using the convolution of the single layer degree distributions.

\begin{align}
P_{L} &\sim P_{k_{\min}}*\overset{\underbrace{L}}{\cdots}*P_{k_{\min}}
\label{eq:convolutions}
\end{align}

The probability that a topological configuration that enables the fulfilment of the payoff conditions specified by Eq.~\eqref{eq:condition_no_invasion} exists, is given by:
\begin{equation}
P_{\mbox{\scriptsize survival}}=\sum_{k_1=k_{\min}}^{\infty}{P_{k_{\min}}(X=k_1)\sum_{q=k_{\min}\cdot L}^{\infty}{P_{L-1}(X=q)\cdot P_{L}{(X\le\lfloor q + k_1\cdot T - S + 1\rfloor )}^{k_{1}}}}
\label{eq:probSurvival}
\end{equation}
where $q$ is the payoff obtained by the defector node playing as a cooperator in $L-1$ layers. With that information, an upper bound for the aggregated degree of the defector's neighbours can be defined as $\lfloor q + k_1\cdot T - S + 1\rfloor$, and if all the neighbours have an aggregated degree below this upper bound, the defector can survive. It is worth noticing that the upper bound for the degree of a cooperator is a discretization of payoff values that involve $S$ and $T$. This means that the survival probability of a defector only changes when the relation between $S$ and $T$ changes by an amount large enough.

The expression for the degree distribution probability function is for an Erd{\H{o}}s-R\'{e}nyi network, assuming that we have a restriction for the minimum degree, so the degree distribution follows a Poisson distribution given by:

\begin{equation}
P_{k_{\min}}(X=k) = \frac{\lambda^{k}\cdot e^{-\lambda}}{k!\cdot(1-e^{-\lambda}\cdot\sum_{i=0}^{\lfloor k_{\min}\rfloor}\frac{\lambda^i}{i!})}
\label{eq:probKminPoiss}
\end{equation}

\begin{figure*}[t!]
    \centering
    \includegraphics[width=0.95\textwidth]{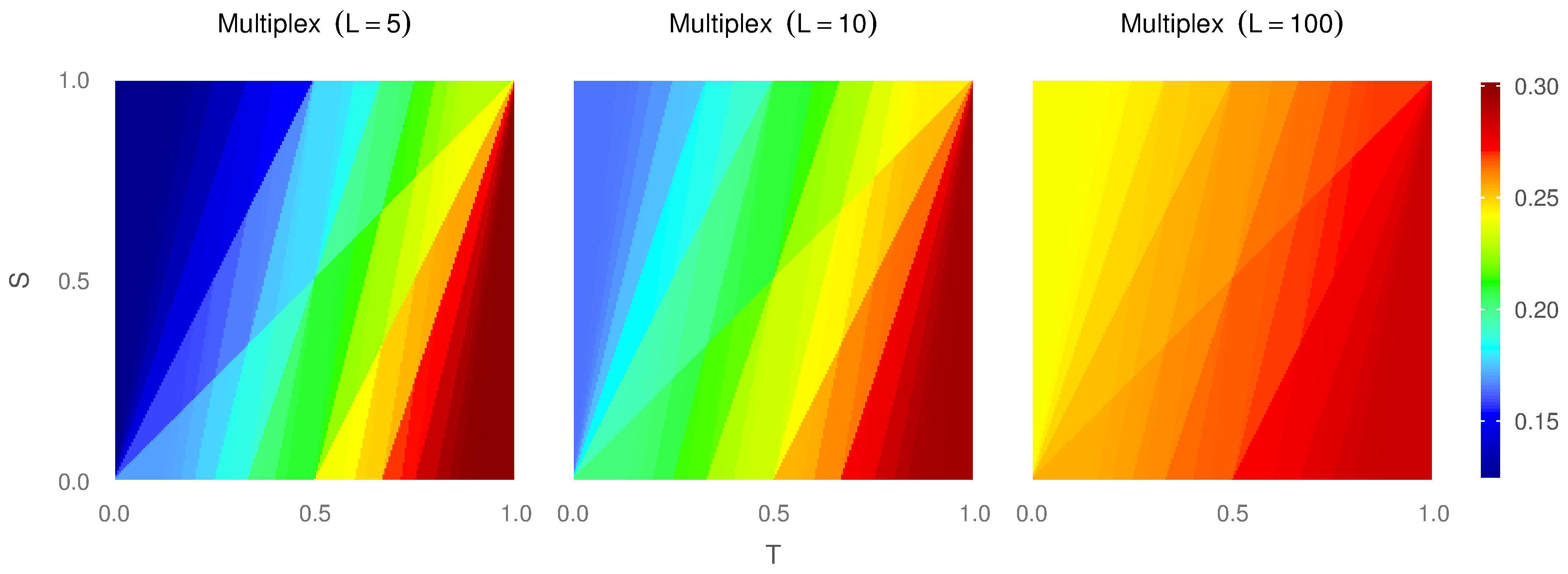}
    \caption{Probability of a defector surviving in the Harmony Game for 5 layers (left), 10 layers (middle) and 100 layers (right), calculated according to Eq.~\eqref{eq:probSurvival}. The individual layers are ER with $\langle k \rangle =3$.}
\label{fig:probHarmony}
\end{figure*}

In Fig.~\ref{fig:probHarmony}, we show the probability of a defector surviving in a full-cooperative population, calculated numerically using Eq.~\eqref{eq:probSurvival}. We observe that this probability increases naturally with $T$, because this is the payoff that a defector obtains against a cooperator, but it is only slightly dependent of the payoff of a cooperator against a defector, $S$. The number of layers has a huge impact on this probability: as the number of layers increases, the probability becomes more uniform in the $T-S$ plane, increasing in general. This can be explained by the relative contribution to the accumulated payoff that comes from layer 1 (the layer where the defector survives): the more layers are added to the system, the smaller this relative contribution. For a large number of layers, this implies that the values of S and T (that determine the payoff)  are less important in the probability of a defector persisting in the system. For networks with a higher mean degree (see Fig.~\ref{fig:prob_survival_defector} in the Supplementary Information), however, the chances of a defector surviving are lower: if the number of neighbours of the defector node is higher, then the probability that one of them has more payoff than him is also higher, thus the defector will tend to imitate the neighbour's behaviour (or in other words, his chances of survival will decrease).

\textbf{Coherent Cooperation.}
Prompted by the topological configurations described earlier, we can now define a ``coherent cooperator'' as a node that, at a given instant of time, plays as cooperator in all $L$ layers of the system. Similarly, we can define a  ``coherent defector'' as a node that, at a given instant of time, plays as defector in all $L$ layers of the system. Finally, those individuals that are neither coherent cooperators nor coherent defectors will be called   ``incoherent'' individuals. This new terms introduced here should not be mistaken for the concepts ``pure cooperators'',  ``pure defectors'' and  ``fluctuating individuals'' introduced in~\cite{GomezGardenes:2006fi}, which implied a \textit{temporal} consistency of the agents' strategies. Also, we want to stress that a incoherent individual as defined here, is clearly different from the concept of a mixed population, that refers simply to a set of both strategies, coexisting together in a population. Moreover, we have to take into account that a coherent behaviour is not trivial nor easily reachable, due to the fact that our simulations start with all mixed populations (randomly distributed and uncorrelated strategies in all layers), so the dynamics that leads to coherence is specially interesting to study.

\begin{figure*}[t!]
    \centering
	\includegraphics[width=0.95\textwidth]{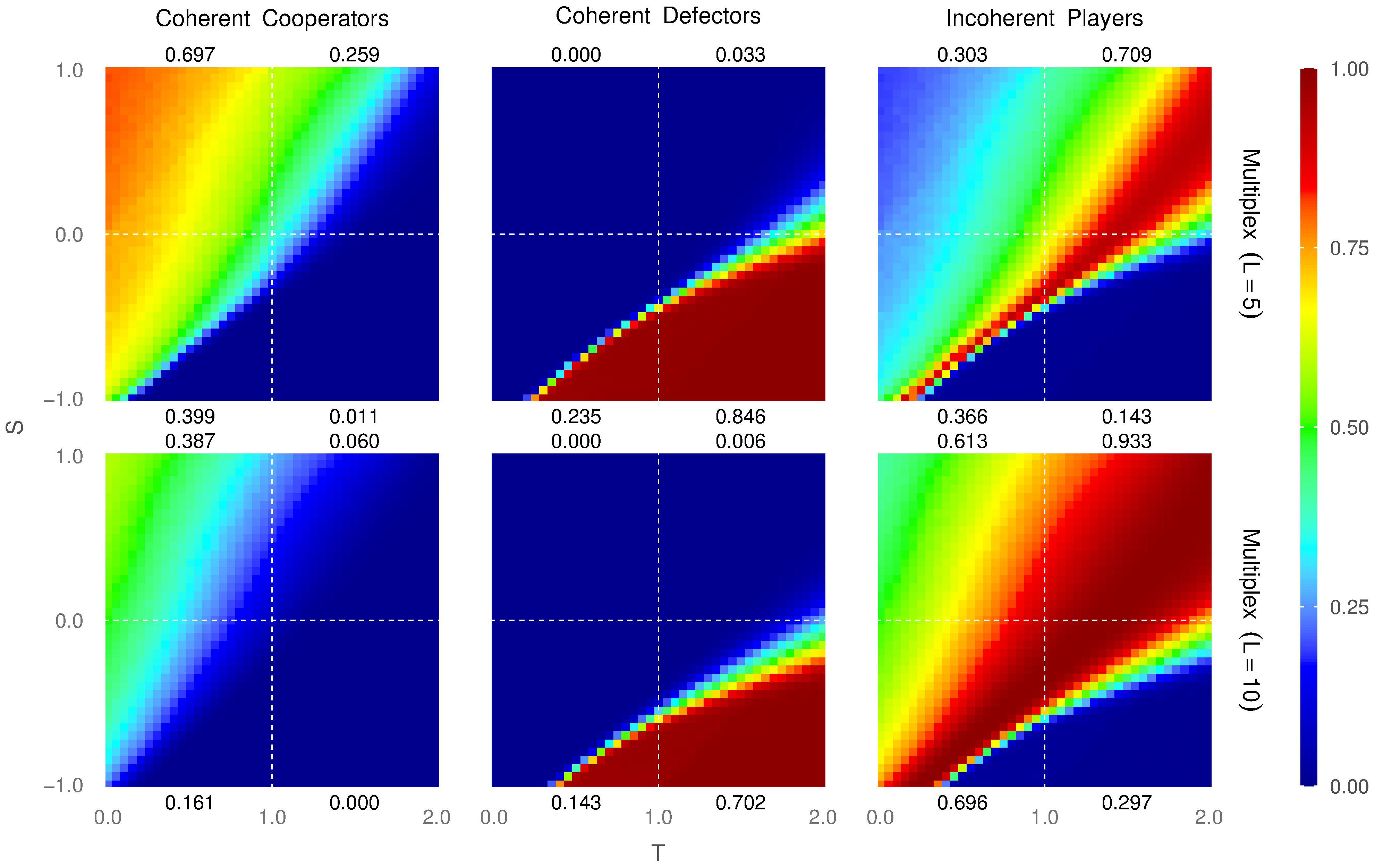}
    \caption{Average density of coherent cooperators (left column), coherent defectors (middle column) and incoherent individuals (right column) for networks with 5 layers (top row) and 10 layers (bottom row). The average density of the corresponding type of individuals is also provided for each one of the quadrants (upper-left is the Harmony Game, upper-right is the Snow Drift, Stag-Hunt is the lower-left, and the Prisoner's Dilemma in the lower-right).}
\label{fig:mega_panel_coherence}
\end{figure*}

In Fig.~\ref{fig:mega_panel_coherence} we show the fraction of coherent cooperators (left column), coherent defectors (middle column) and incoherent individuals (right column) for 5 layers (top row) and 10 layers (bottom row). The formation of coherent cooperators is particularly complicated, and it is interesting to notice that even in the Harmony game there is a low fraction of them (except for a small area around the extreme case of $T=0$ and $S=1$). In the other quadrants, the fraction is very small (in particular, the Prisoner's Dilemma presents basically no coherent cooperation). This implies that most of the cooperation shown by the system comes from incoherent individuals. We also observe that the fraction of coherent cooperators decreases quickly with the number of layers for any game. As we have said, the origin of such results resides in the fact that a defector takes advantage of its own cooperative behaviour in other layers, specially in regions of the $T-S$ plane prone to cooperation.

Conversely, regarding the fraction of coherent defectors, we observe that their presence is very strong in most of the Prisoner's Dilemma region and part of the Stag-Hunt area, and they decrease only slightly when increasing the number of layers from 5 to 10. This fact is easy to understand: the resilience of a cooperator in a hostile environment is based basically in how he performs as cooperator, the advantage of playing as defector in other layers is practically zero because in a large defector population the contribution to the payoff of a defector that plays against a defector is zero, $P=0$. Thus, in these regions, the survival rate of cooperation does not improve by playing as defector strategy in other layers.

Regarding incoherent individuals, we observe that they are very prevalent for all games (except for the extreme area of Harmony around ($T=0, S=1$), where cooperation is very profitable, and the bottom-half area of the hard Prisoner's Dilemma where cooperation is extremely expensive).  Incoherent individuals contribute significantly to the average density of cooperation in a large central area of the $T-S$ plane, particularly in the areas that separate full-cooperation from full-defection (See also Fig.~\ref{fig:mega_payoff} in the Supplementary Information for a detailed description of the fraction of incoherent individuals playing as cooperators). This area of prevalent incoherent individuals increases with the number of layers or, in other words, it gets harder and harder to be a coherent strategist as the number of layers increases.

Fig.~\ref{fig:mega_panel_coherence} also confirms what we showed analytically earlier: defection can survive in the Harmony game, as long as the individual defecting in a particular layer is a incoherent individual; it plays as cooperator in other layers and obtains enough payoff from them to avoid having to switch strategies (see also Fig.~\ref{fig:mega_payoff} in Supplementary Information for further detail on the payoff of cooperators and defectors).

Interestingly enough, in Fig.~\ref{fig:mega_panel_coherence} we can observe how coherent players of opposite types do not coexist in the same population. Another important point is where coherent players can coexist with incoherent players. The area where coherent cooperators interact with incoherent players is wide and gets wider as we keep adding layers to the multiplex. However, the area of coexistence of coherent defectors and incoherent players is very narrowed and is only slightly affected when layers are added to the structure. This means that the coherent defection is a very dominant strategy that almost forbids the existence of any other kind of players.

\section{Discussion}
\label{sec:discussion}

In this paper we have presented a systematic and comprehensive analysis of the outcomes of cooperation dynamics on ER multiplex networks for the four games on the $T-S$ plane, when using the Replicator updating rule, comparing our results with those already known for the case of the games on monoplex. Also, we have analyzed the microscopic behavior of the nodes, and coined the terms of coherent cooperator, coherent defector and incoherent player.

In particular, we have found that the stationary distribution of cooperation in the plane $T-S$ becomes less sharp as more layers are added. In the monoplex case there is a very narrow area that separates all-cooperator from all-defector areas for the Stag Hunt and Prisoner's Dilemma games, but in the multiplex scenario we find that it becomes a wider region, with intermediate values of cooperation. We also find that the region of all-defectors shrinks as the number of layers increases. As a counter-effect though, we find a slight decrease in the value of cooperation (even in the quadrant of the Harmony game), from total cooperation to values around $90\%$. These results are consistent with and generalize those found by~\cite{GomezGardenes:2012hca}: the introduction of a multiplex structure in the population helps promote cooperation in regions of the parameter space in which it can not survive in the monoplex scenario, at the expense of a moderate decrease of cooperation in those where traditionally it was very high. We explored the microscopic underpinnings for these phenomena, previously observed but unexplained in the aforementioned paper.

Thus, regarding the microscopic behavior of the nodes, we have found that in general and at a given time step, there are three types of individuals: those coherently acting as cooperators in all layers, those acting as coherent defectors, and a group of incoherent individuals, that play as cooperators in some layers and as defectors in others. The existence of this third incoherent group is at the root of the explanation of the survival of defection in the Harmony Game for a multilayered network, and it is also responsible for a large part of the cooperation in the central areas of the $T-S$ plane, where cooperation is lower in a monoplex. Also, we have analyzed how this three types of players interact among them, concluding that there are plenty of interaction between incoherent and coherent cooperators, fewer  interactions between incoherent and coherent cooperators, and practically no interaction between both types of coherent players. Moreover, this is a very plausible social scenario: some people may behave consistently in all their types of interactions (for example at work, at home, with friends,etc) either cooperating or defecting, and some other may choose different strategies for different layers (for example, cooperate with family and defect at work). We have found that an the fraction of incoherent players increases with the number of layers increases, which means that as the number of contexts where the a players interact increases, it gets harder to maintain a coherence  behaviour in all of them. Regarding the dependence with the initial fraction of cooperation, we found that our system behaves consistently with what was found for the monoplex network, and the effect of adding more layers is preserved or even increased with increasing initial fraction of cooperators.

To summarize, the introduction of multiplex networks not only is a more realistic representation of social systems, allowing for more sophisticated individual behaviours, but as it has been shown in other context too, it has a profound effect on the dynamics developing on top of them.

{\bf Competing financial interests:} The authors declare no competing financial interests.

{\bf Author contributions:}  J.T.M. performed simulations and prepared figures. J.T.M., S.G., A.A. and J.P.C. contributed equally to the design of the experiments, interpretation and discussion of the results, and writing of this paper.

{\bf Acknowledgements:} Authors acknowledge support from MINECO
through Grant FIS2012-38266; the EC FET-Proactive Project MULTIPLEX (grant 317532), ICREA Academia and the James S.\ McDonnell Foundation.

\pagebreak
\widetext
\begin{center}
\textbf{\large Supplementary Information: Strategical incoherence regulates cooperation in social dilemmas on multiplex networks}
\end{center}
\setcounter{equation}{0}
\setcounter{figure}{0}
\setcounter{table}{0}
\setcounter{section}{0}
\setcounter{page}{1}
\makeatletter
\renewcommand{\thesection}{S\arabic{section}}
\renewcommand{\theequation}{S\arabic{equation}}
\renewcommand{\thefigure}{S\arabic{figure}}
\renewcommand{\bibnumfmt}[1]{[S#1]}
\renewcommand{\citenumfont}[1]{S#1}
\section{Convergence}
\label{sec:convergence}

\begin{figure*}[t!]
	    \centering
	    \includegraphics[width=1\textwidth]{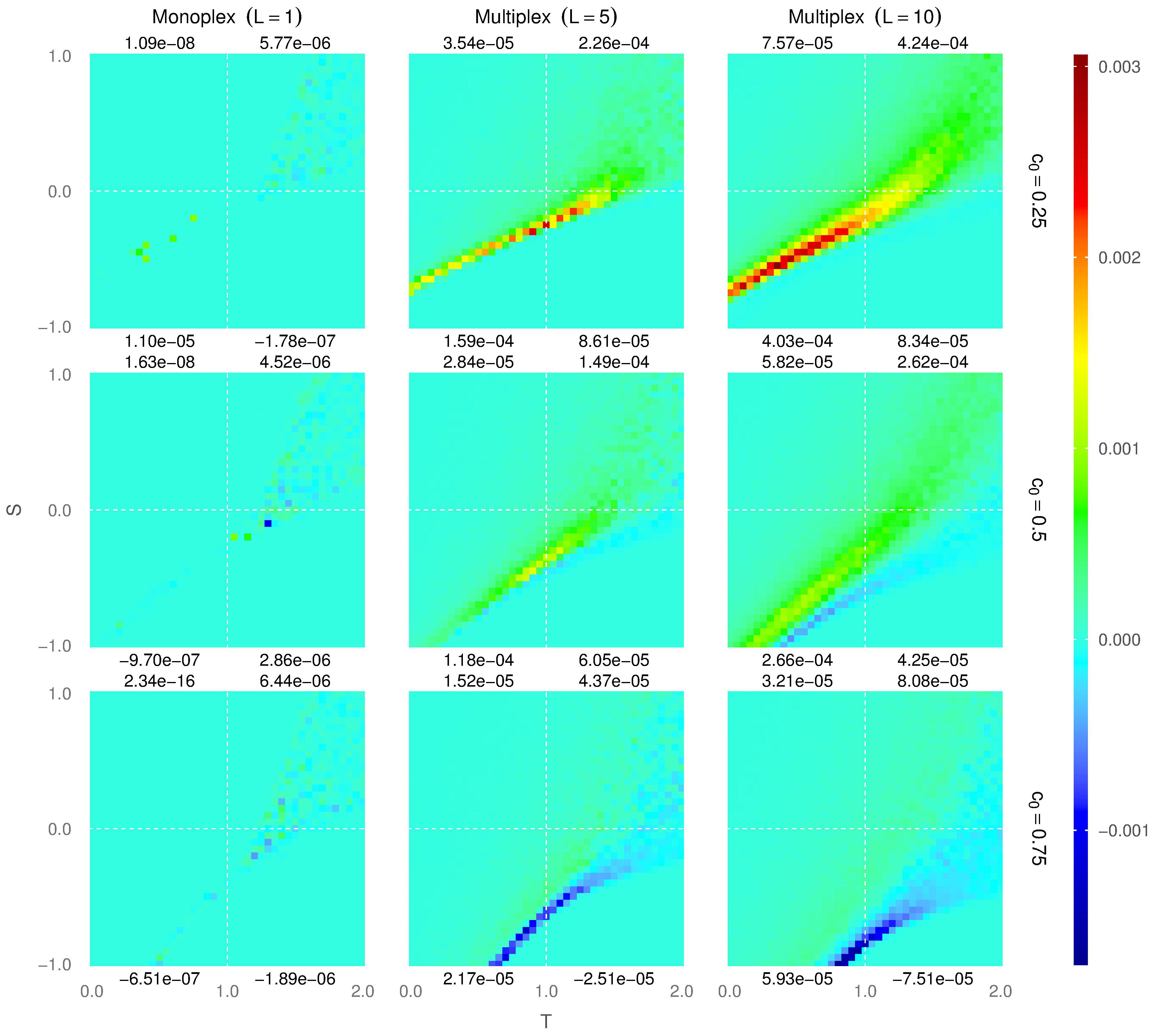}
	    \caption{Convergence to the stationary state measured as the average variation in the fraction of cooperators during 1000 time steps, measured using the slope of a linear model fitted at the end of $t_{\gamma}$ steps of the simulation. The numbers in each quadrant represent the mean convergence value (the slope of the fit) for each one of the four games (upper-left is the Harmony Game, upper-right is the Snow Drift, Stag-Hunt is the lower-left, and the Prisoner's Dilemma in the lower-right). In the different rows we show the information for several values of initial fraction of cooperators ($c_0=0.25$, $c_0=0.50$, $c_0=0.75$), while the different columns correspond to 1, 5 and 10~layers, respectively.}
	    \label{fig:convergence}
\end{figure*}

We study the system's convergence to the stationary state. It is well known that, in general, the time evolution of cooperation on a monoplex network, for a value of the parameters that allows the survival of at least some cooperation, usually follows a curve that initially decreases moderately, while cooperation rearranges itself from the random initial conditions into a more favorable setting (either in one or multiple clusters), and then there is a new increase, followed by the achievement of the stationary state. In general, this whole transient time is relatively short (typically of $1-2 \times 10^{4}$) for a size of $1-4 \times 10^{3}$ nodes. However, it hasn't been explored in detail until now the convergence process for the four games in the $T-S$ plane on multiplex networks.

In order to evaluate such convergence, we fit the last $t_{\gamma}$ time steps of the evolution to a linear trend, $\widehat{c}(t) = \alpha + \beta t$ using the QR decomposition method. Then we use the slope of the fitted model to compute the variation of the density of cooperators every 1000 time steps, $\Xi=1000\cdot\beta$. Thus, a near-zero value of this metrics indicates that the system has reached the stationary state, while a positive value would indicate that the average level of cooperation is still increasing in the system at that time, and vice versa. Figure~\ref{fig:convergence} shows how every point of the $T-S$ plane performs on our measure of convergence during the last $t_{\gamma}$ time steps of the simulation.

Monoplex networks (left column in Figure~\ref{fig:convergence}) seem to reach the stationary state according to our convergence criteria for every point of the plane $T-S$ and independently of the initial fraction of cooperators: the slope is in general smaller than $10^{-4}$. We observe, however, a small amount of stochastic noise for some regions of Snow-Drift and Prisoner's Dilemma games, where our measure indicates that the stationary is not fully reached. Nonetheless, we will show in Section~\ref{sec:fluctuations} that this noise is just an effect of the large fluctuations in the number of cooperators when the stationary is reached.

In multiplex networks, on the other hand, there is a non-negligible area where convergence is not reached (red areas in the central and right panels in the first row in  Figure~\ref{fig:convergence}). In the most extreme cases, where the slope of the linear model $\beta$ is largest, our measurements indicates an increment of the cooperators of about a 0.1\% every 1000 time steps. That could seem a smaill increase in the fraction of cooperators, however if the evolutionary process were to run for a very long period of time, the increase could be significant.

\begin{figure*}[t!]
    \centering
    \includegraphics[width=0.47\textwidth]{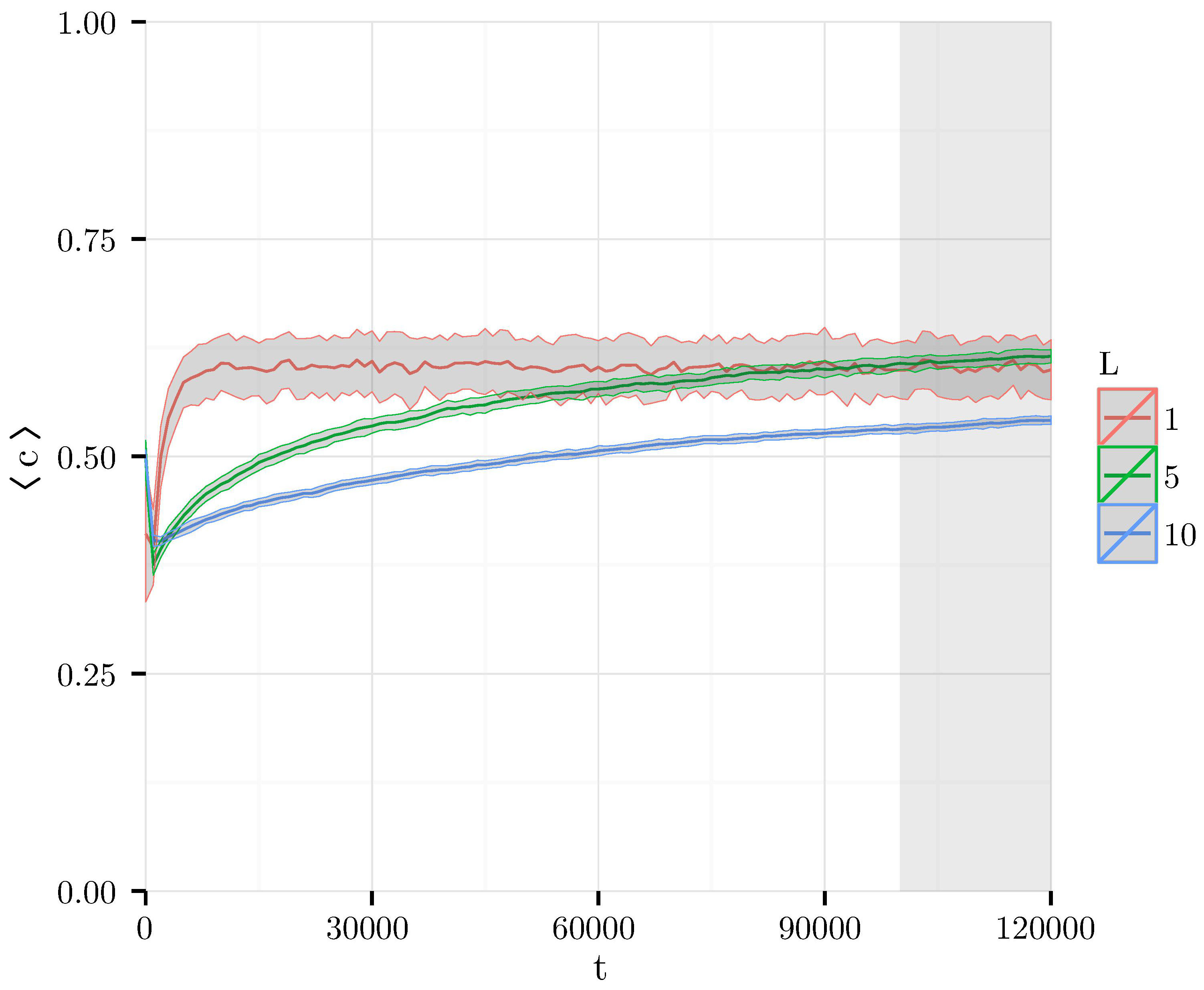}
    \caption{Example of the time evolution of cooperation for the point in the plane $T-S$ with maximum fluctuations values ($S = 0$ and $T = 1.35$). The shadowed area for each plot represents two standard deviations over the residuals of the $I$ iterations at each time step. It will be used to compare the size of the fluctuations between monoplex and multiplex networks. The grey vertical area corresponds to the interval $[t_0, t_0+t_{\gamma}]$ where the measures shown in all panel figures in this paper are computed.}
    \label{fig:evofluctuations}
\end{figure*}

To better illustrate this difference in the path to stability for monoplex vs. multiplex networks, we show in  Figure~\ref{fig:evofluctuations} the time evolution of the level of cooperation, $\langle c \rangle$ for a single simulation progresses (monoplex plotted in red and two multiplex networks with different number of layers represented in green and blue), for one point in the plane $T-S$. This particular point has been picked as an extreme case, for having the maximum fluctuation values in the entire $T-S$ plane (see Figure~\ref{fig:fluctuations} and Section~\ref{sec:fluctuations} for further detail). We clearly observe that, while the time required for the monoplex system to achieve the stationary state is around $1-2 \times 10^{4}$, for the multiplex networks it can be at least one order of magnitude larger, and it increases with the number of layers, too. However, it is important to remember that this example shown here is a very extreme case, while the convergence process in multiplex is in general faster for regions of the plane that are far away from the transition area (or areas where the final state is close to an all-cooperation or all-defection).

To understand the reason for such an increase in the convergence time for multiplex with respect to monoplex (at least in some regions of the $T-S$ plane), one has to pay attention to which areas are more reluctant to reach stability. Such regions correspond again to the transition areas between those that end up in total cooperation and those that end up in total defection. In the Stag Hunt quadrant, the game has an unstable equilibrium with mixed population, which means that the game will tend to converge to total cooperation or total defection as happens in the monoplex network. However the multiplex structure of the network changes that outcome, as we described in the Results Section. In these structures, the transition region is larger than in the monoplex, and is in this transition region where the convergence is hard. The analysis of how the fraction of cooperators has an effect on the convergence gives us an insight about what is happening. We have already stated that the interlayer dynamics has an important role in the survival rate of defectors and cooperators. If we look at the multiplex columns of the Figure~\ref{fig:convergence}, we can observe how the convergence is strongly affected by the initial fraction of cooperators. On the one hand, if the initial fraction of cooperators is small, they will need more time to reach the equilibrium because they have to fight against a larger fraction of defectors that benefits from the interlayer dynamics. On the other hand, a larger initial number of cooperators implies that the defectors will need more time to reach an stable configuration. However, the presence of a large number of initial cooperators has less impact on the convergence; which is easily understood, given the fact that defectors get more profit from cooperating in other layers than the opposite case.

Similar conclusions could be reached for the other games, taking into account that the transition regions between full-cooperation and full-defection are different in nature, for instance in the Snow-Drift this region is wider. Thus, we can see the effect of non-convergence is diluted across the Snow-Drift quadrant.

\section{Analysis of fluctuations}
\label{sec:fluctuations}

We turn our attention now to the fluctuations of the system in the stationary state. In the case of these four games on a monoplex network, it is well known that the level of cooperation in the stationary state fluctuates around a well-defined average value due to the effect of both the topological structure of the network and the nature of the Replicator updating rule. We propose a measure in order to quantify these fluctuations and later compare them with the cases of multiplex networks. For each one of the $I$ repetitions of the experiment we fit a linear model to the final $t_{\gamma}$ time steps of the simulation, $\widehat{c_i}(t) = \alpha_i + \beta_i t$. We also need to take into account a possible non-zero slope in the measure of fluctuations (see in Section~\ref{fig:convergence}), so we average the Mean Square Error between the data from the simulations and the predictions of the linear model for the $I$ iterations, calculated as:

\begin{equation}
    \zeta = \frac{L}{I}\sum_{i=1}^{I}{MSE_i} = \frac{L}{I\cdot t_{\gamma}}\sum_{i=1}^{I}{\sum_{t =t_0}^{t_{\gamma}+t_0}{{(c_i(t) - \widehat{c_i}(t))}^2}}.
    \label{eq:fluctuations}
\end{equation}

\begin{figure*}[t!]
	    \centering
	    \includegraphics[width=1\textwidth]{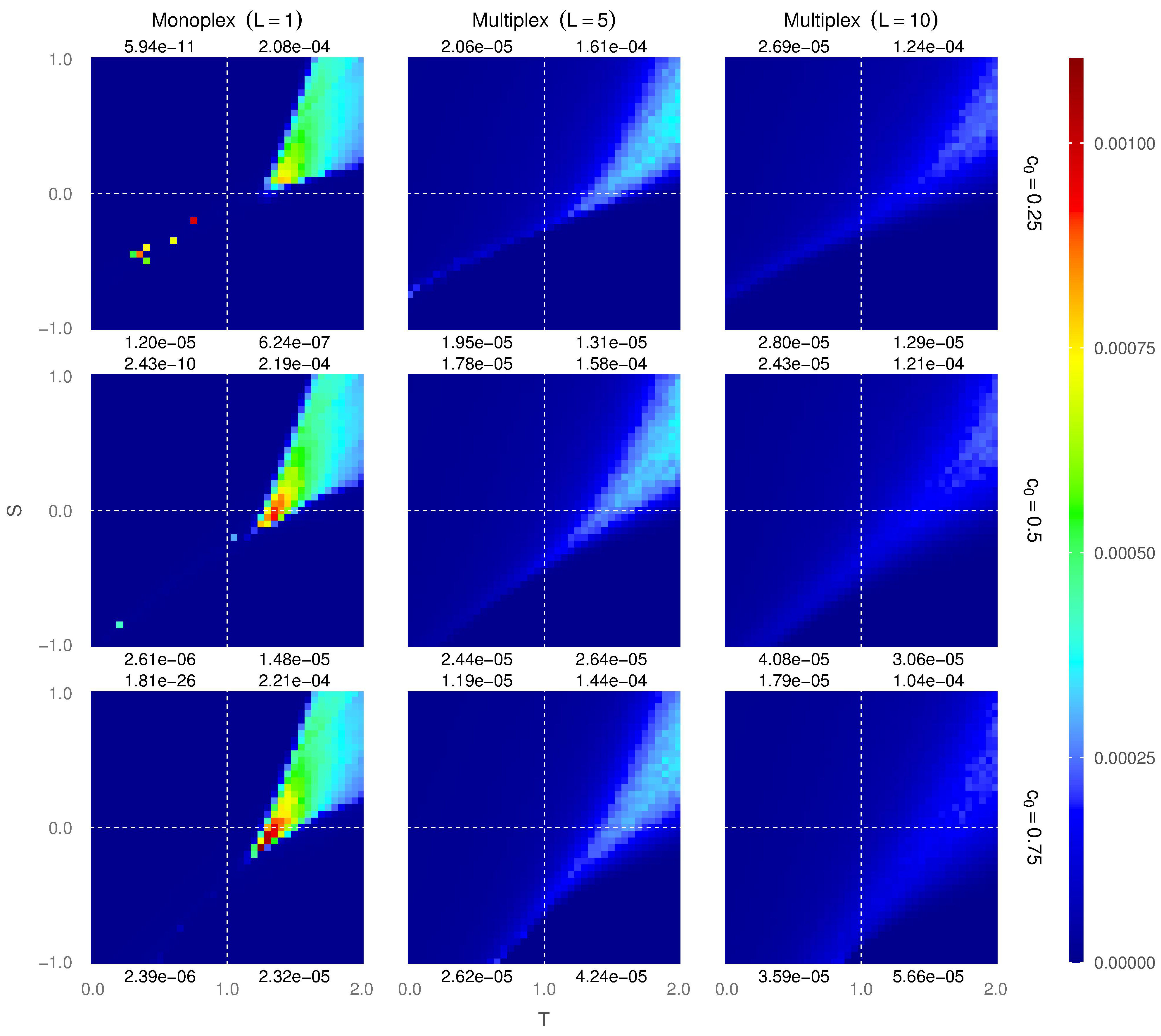}
	    \caption{Fluctuation of the fraction of cooperator around the fitted trend at the final time steps of the simulation computed as described in the equation~\eqref{eq:fluctuations}, for each pair of values S and T. The number in each quadrant represents the average value for each one of the four games (upper-left is the Harmony Game, upper-right is the Snow Drift, Stag-Hunt is the lower-left, and the Prisoner's Dilemma in the lower-right). The results are provided by 3 different initial conditions}
	    \label{fig:fluctuations}
\end{figure*}

The results for monoplex and multiplex networks are displayed in Figure~\ref{fig:fluctuations}. For the monoplex case, the simulations show small fluctuations in the quadrants of Stag-Hunt and Harmony Game. However for the Snow-drift on monoplex, the results display a zone where fluctuations are larger than in the rest of the plane $T-S$. That can be attributed to the nature of the game: it has an evolutionary stable equilibrium with mixed populations, so a consensus where both strategies coexist has to be reached. To achieve this objective some nodes have to alternate their strategies. These changes, due the topological features of the network, can lead to a cascade effect of changes in a large portion of the network; the equilibrium gets disturbed, and a new  equilibrium has to be reached again. This causes the relatively large fluctuations that we  measure. It is worth noticing that in the area of mutual coexistence of strategies the fluctuations are larger where the temptation and sucker payoffs are not far from the payoffs of mutual cooperation and mutual defection. It is also noteworthy that, even when the Prisoner's Dilemma quadrant presents very small fluctuations in general, it does show a small but very significant spot near the line of weak Prisoner's Dilemma, where they are large. Again, this corresponds to the area of competition between Cooperation and Defection, where each of the strategies accounts roughly for half the population.

The introduction of multiplex networks has an enormous effect on the fluctuations. The fluctuations are again in the region of coexistence of strategies, however, in the case of 5-layer multiplex the fluctuations are much smaller than in the monoplex case. The results in the 10-layer multiplex display an even larger reduction in the measure of fluctuations (compare also the three example curves shown in Figure~\ref{fig:evofluctuations}). The nature of such reduction from monoplex to multiplex is to be found in the interlayer dynamics. Each layer is pushed to reach an stable equilibrium where both strategies can coexist, nonetheless the shared information between the layers establishes a way to constrict the range of the fluctuations. The change of strategy of a node in one layers is not conditioned by its performance in that single layer, but by its global performance in the entire multiplex structure. That makes the system more robust to fluctuating nodes, at the expense of convergence time to the stationary equilibrium. We observe that, both for monoplex and multiplex structures, the initial fraction of cooperators, $c_0$, barely has any observable influence on the size of the fluctuations.

It is worth mentioning that the fluctuations shown in Figure~\ref{fig:evofluctuations} are calculated with a modified version of equation~\eqref{eq:fluctuations}, as follows: we have to divide the time range in different slices in the interest of realizing local accurate measures of the fluctuations in each slice, so we fix the size of the time window to $t_w = 1000$. For each time slice we fit a linear model to each of the $I$ runs of our simulation, then we compute the residuals as the difference between the linear model and the data from the simulations. We have $I$ residuals that measure the size of the fluctuations at each time step, and we plot a range corresponding to twice their standard deviation to provide information about the size of the fluctuations at each time step.

\section{Probability of a defector surviving in the Harmony Game for higher average degree}
\label{sec:prob_survival_defector}

\begin{figure*}[t!]
	    \centering
	    \includegraphics[width=1\textwidth]{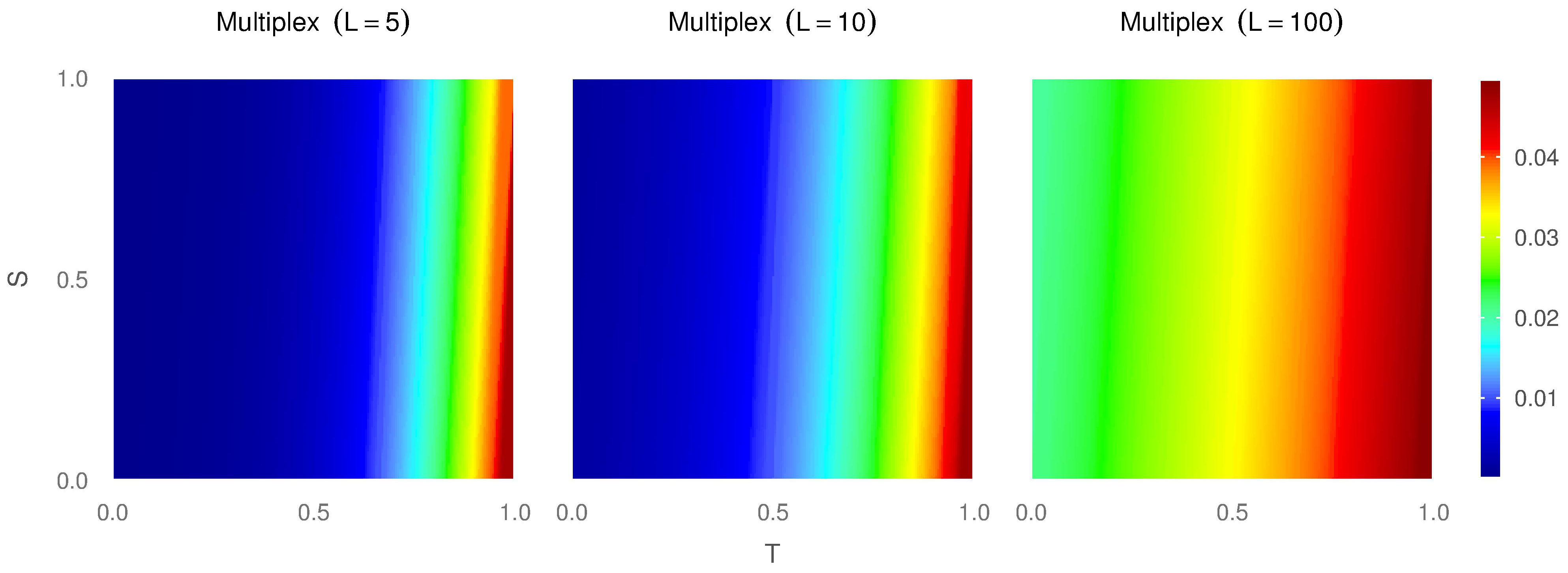}
	    \caption{Probability of a defector surviving in the Harmony Game for 5 layers (left), 10 layers (middle) and 100 layers (right) for an average connectivity of $\langle k \rangle =20$, calculated according to equation~15 from the main text.}
\label{fig:prob_survival_defector}
\end{figure*}

In Figure~\ref{fig:prob_survival_defector} we show the probability of a defector surviving in a full-cooperative population, calculated numerically using equation~15, for the case of a higher average degree, $\langle k \rangle =20$, than in the main text. The main impact of an increased value of average degree is a significant decrease of the probability, for any number of layers, or values of $S$ and $T$. The effects discussed in the main text remain for this case too, but attenuated (note that the range of values for the probability are smaller in this case). In general, the probability increases with $T$, but it is only slightly dependent of $S$. As the number of layers increases, the probability becomes more uniform in the $S-T$ plane, increasing in general.

\section{Percentage of cooperation among mixed individuals, and payoff of cooperators and defectors}
\label{sec:percentage_coop}

In this final section, we address the analysis of the percentage of cooperation among mixed individuals, as well as the payoff obtained by both cooperators and defectors.
Regarding the former, we observe that the percentage of mixed individuals playing as cooperators is very high in the Harmony game, and in the upper diagonal of the Stag-Hunt game, as well as the upper diagonal of the Snow-Drift. In the Prisoner's Dilemma game, however, it is zero except for a small region near the weak limit, when cooperation is relatively inexpensive.  This general situation gets emphasize by the increasing of the number of layers.

\begin{figure*}[t!]
	\centering
	\includegraphics[width=1\textwidth]{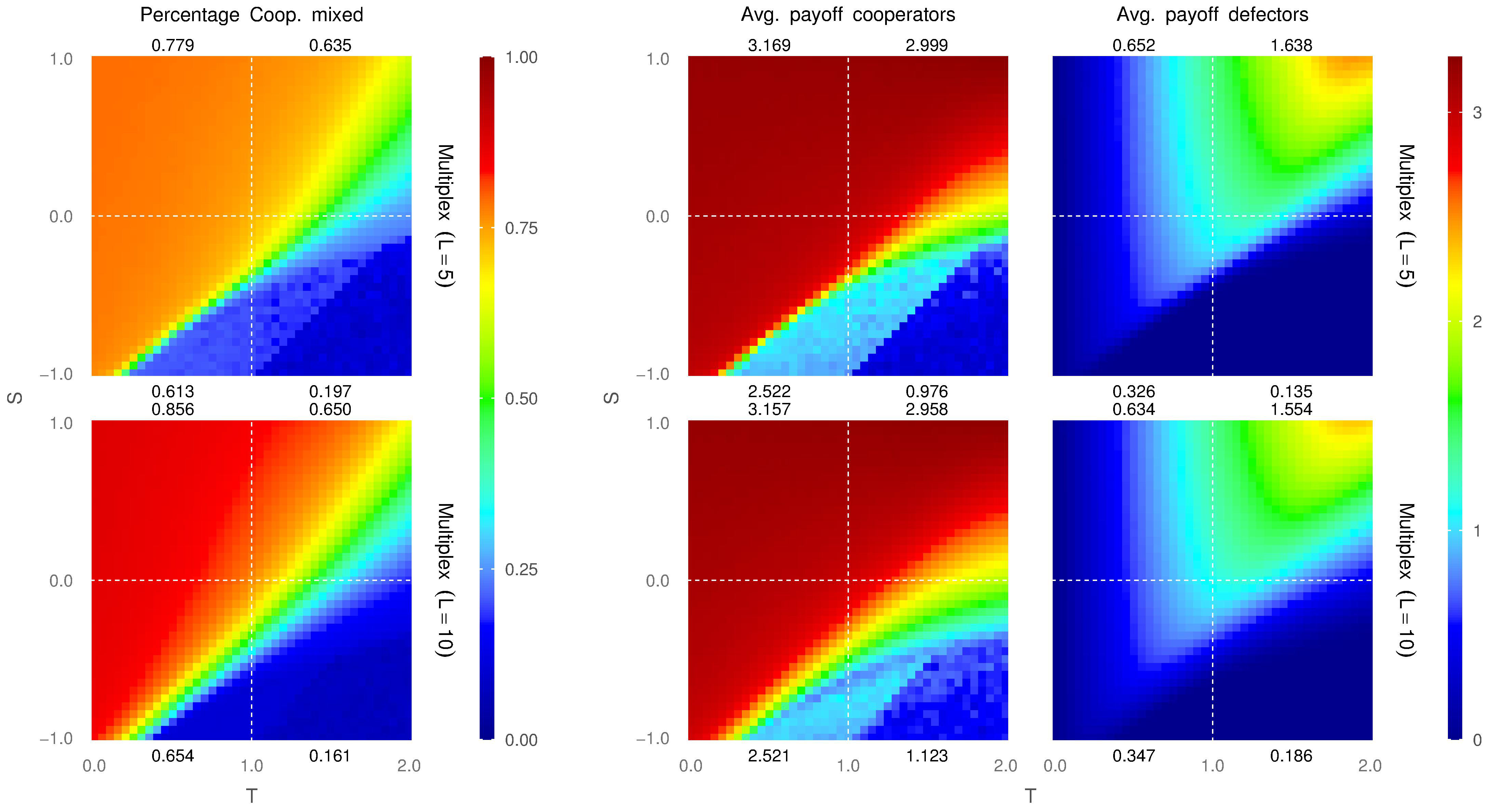}
	\caption{Percentage of cooperation in mixed individuals (left column), average direct payoff obtained playing as cooperator (middle column) and average direct payoff obtained playing as defector (left column) for 5 layers (top row) and 10 layers (bottom row) multiplex, in the four games. The corresponding averaged values over the quadrants are also provided (upper-left is the Harmony Game, upper-right is the Snow Drift, Stag-Hunt is the lower-left, and the Prisoner's Dilemma in the lower-right).}
\label{fig:mega_payoff}
\end{figure*}

Regarding the latter, we observe that the average payoff among cooperators in obviously the highest in the Harmony game, and upper diagonals of both Stag-Hung and Snow-Drift, and it is zero in the hardest, bottom diagonal of the Prisoner's Dilemma game, with a wide transition area of intermediate values separating both regions. Moreover, this description seems to be independent of the number of layers in the system. Finally, the only regions where defection gets a moderate payoff are within the Snow Drift game, while it is near zero anywhere else. This picture is also independent of the number of layers.

\section{Initial fraction of cooperators.}
\label{sec:initial_conditions}
\begin{figure*}[t!]
    \centering
    \includegraphics{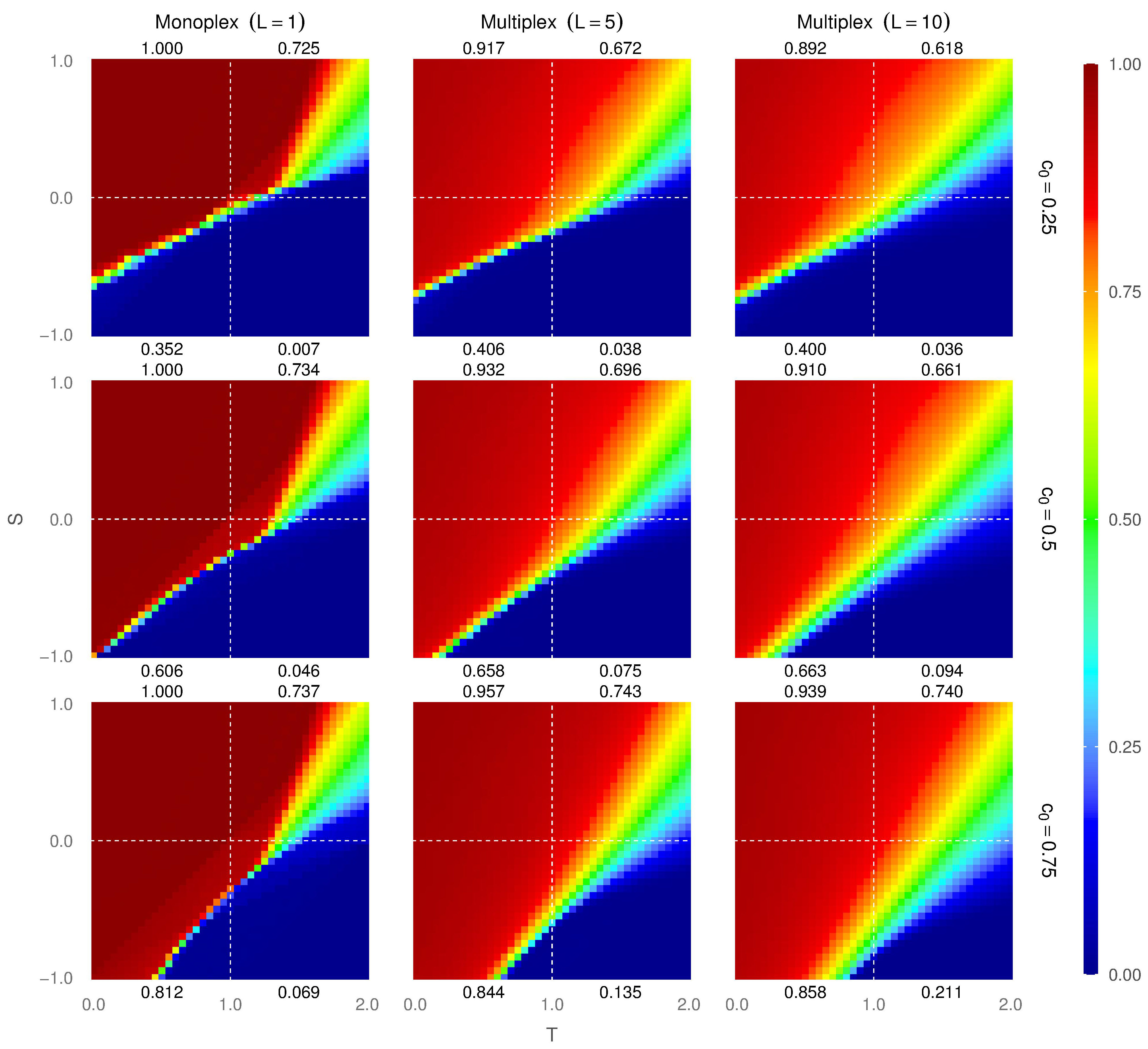}
    \caption{Asymptotic density of cooperators $\langle c\rangle$ for networks with different number of layers ($L=1$ in the left column, $L=5$ in the central column, $L=10$ in the right column), and different initial fraction of cooperation ($c_0=0.25$ in the top row, $c_0=0.5$ in the central row, $c_0=0.75$ in the bottom row). The plane $T-S$ is divided into four major regions that correspond to the four games under study: the upper-left area is the Harmony Game, the upper-right is the Snow Drift, Stag-Hunt is in the lower-left, and the Prisoner's Dilemma is in the lower-right. The average asymptotic density of cooperators for each one of the games is also indicated, as a numerical value, next to the corresponding quadrant.}
\label{fig:initialcond}
\end{figure*}

For completeness, in Fig.~\ref{fig:initialcond} we show the stationary average fraction of cooperation for the four-game plane and various numbers of layers, for three different initial fractions of cooperators (upper row is $c_0 =0.25$, middle row  is $c_0 =0.5$ and bottom row is $c_0 =0.75$). We will briefly discuss now the differences between the previously explained case of $c_0 =0.5$, and the other two scenarios.

We observe that, for a given game quadrant and a given number of layers, increasing the initial fraction of cooperation has in general a positive but moderate impact on the stationary fraction of cooperation, specially in the Stag Hunt and Prisoner's Dilemma games. In the former one, we have an unstable evolutionary equilibrium in mixed populations, so the change of $c_0$ has a significant impact on the final outcome. In the case of Prisoner's dilemma game, an increase in the initial fraction of cooperators means an increase in the probability that clusters of cooperators forms.

The effect discussed in this paper when adding layers to the system still holds or is even emphasized by an increased initial fraction of cooperators: the overall stationary value of cooperation increases with the number of layers in the Prisoner's Dilemma Game and Stag-Hunt, the region of coexistence between both strategies widens for the Snow-Drift Game, and the Harmony game presents a small decrease of cooperation.

\end{document}